\documentstyle[eqsecnum,prd,aps,epsf]{revtex}
\draft
\begin{document}
\newtheorem{th}{Theorem}
\newtheorem{pr}{Proposition}
\newtheorem{lm}{Lemma}

\thispagestyle{empty}
{\baselineskip0pt
\leftline{\large\baselineskip16pt\sl\vbox to0pt{\hbox{\it Department of Physics}
               \hbox{\it Kyoto University}\vss}}
\rightline{\large\baselineskip16pt\rm\vbox to20pt{\hbox{KUNS-1551}
            \hbox{\today} 
\vss}}%
}
\vskip1cm
\begin{center}{\large \bf
Nakedness and curvature strength 
of shell-focusing singularity in the spherically
symmetric space-time with vanishing radial pressure}
\end{center}
\vskip1cm
\begin{center}
 {\large 
Tomohiro Harada 
\footnote{ Electronic address: harada@tap.scphys.kyoto-u.ac.jp},
Ken-ichi Nakao
\footnote{ Electronic address: nakao@tap.scphys.kyoto-u.ac.jp},
and Hideo Iguchi
\footnote{ Electronic address: iguchi@tap.scphys.kyoto-u.ac.jp}} \\
{\em Department of Physics,~Kyoto University,} 
{\em Kyoto 606-8502,~Japan}\\
\end{center}

\begin{abstract}
It was recently shown that
the metric functions
which describe a spherically symmetric space-time
with vanishing radial pressure
can be explicitly integrated.
We investigate
the nakedness and curvature strength of the
shell-focusing singularity
in that space-time.
If the singularity is naked,
the relation between the circumferential radius and 
the Misner-Sharp mass is given by
$R\approx 2y_{0} m^{\beta}$ with $ 1/3<\beta\le 1$
along the first radial 
null geodesic from the singularity.
The $\beta$ is 
closely related to the curvature strength of
the naked singularity.
For example, for the outgoing or ingoing null geodesic,
if the strong curvature condition (SCC) by Tipler
holds, then $\beta$ must be equal to 1.
We define the ``gravity dominance condition'' (GDC) 
for a geodesic.
If GDC is satisfied for the null geodesic,
both SCC
and the limiting 
focusing condition (LFC) by Kr\'olak hold
for $\beta=1$ and $y_{0}\ne 1$,
not SCC but only LFC holds for $1/2\le \beta <1$,
and neither holds for $1/3<\beta <1/2$,
for the null geodesic.
On the other hand, 
if GDC is satisfied for the timelike geodesic $r=0$,
both SCC and LFC are satisfied for the timelike geodesic,
irrespective of the value of $\beta$.
Several examples are also discussed. 

\end{abstract}

\pacs{PACS numbers: 04.20.Dw}

\section{introduction}
The cosmic censorship hypothesis (CCH) is one of the
most important open problems in classical gravity
(Penrose 1979).
The CCH roughly says that the physically reasonable space-time
contains no naked singularity. 
Since the CCH asserts the future
predictability of the space-time,
it is so helpful that
several theorems on black holes 
have been proved under the assumption of CCH 
(Hawking and Ellis 1973). 
In spite of all the effort,
the censorship has not yet been proved.
In fact, there have been discovered several solutions which
have naked singularities with matter content that
satisfies energy conditions.
Then the curvature strength of singularities
was defined in a hope that weak convergence
would reveal the extendibility of the space-time
in a distributional sense.
In this context, Tipler (1977) defined the strong curvature
condition (SCC), while Kr\'olak (1987) defined weaker condition,
which we call the limiting focusing condition (LFC).

One of the most important examples 
which have naked singularities is the 
Lema\^{\i}tre-Tolman-Bondi (LTB) solution.
This solution describes the spherical collapse of
an inhomogeneous dust ball.
It has been proved that this solution has
naked singularities from generic initial data.
The naked singularities in this solution
are classified to ``shell-crossing'' 
(Yodzis, Seifert and M\"uller zum Hagen 1973)
and ``shell-focusing'' (Eardley and Smarr 1979, 
Christodoulou 1984) singularities.
Newman (1986) showed that,
for a null geodesic from the shell-crossing singularity, 
neither SCC nor LFC is satisfied.
It is widely believed that 
the shell-crossing singularities would be harmless
because they would be dealt with in some distributional sense. 
Newman (1986) also showed that, 
for a null geodesic from the shell-focusing 
singularity which results from generic smooth initial data, 
not SCC but LFC is satisfied.
Hence, shell-focusing singularities will be
more serious to CCH than shell-crossing singularities.
It is important that the strength of the singularity 
is determined by the curvature divergence not only on 
the null geodesic but also 
on the timelike geodesic. 
Recently, Deshingkar, Joshi and Dwivedi (1999) showed that
both SCC and LFC are satisfied for the timelike geodesics
which terminate at the shell-focusing singularity.

It might be thought that,
since the dust is a pressureless fluid,
there appears naked singularity which satisfies LFC.
As an extension of the LTB solution,
we will consider a spherically symmetric space-time
with a fluid that has only tangential pressure.
Magli (1997, 1998) solved an explicit
solution with the mass-area coordinates.
Here we give a formalism to examine the existence of
naked central singularity and its curvature strength.
Next we proceed further by defining 
the ``gravity dominance condition''.
After that we discuss several examples. 

We follow the sign conventions of the textbook by 
Misner, Thorne and Wheeler (1973) 
about the metric, Riemann and Einstein tensors.
We use the units with $c=G=1$.

\section{metric functions and occurrence of naked singularity}
In a spherically symmetric space-time,
the line element is written
in the diagonal form as
\begin{equation}
  ds^2=-e^{2\nu(t,r)}dt^2+e^{2\lambda(t,r)}dr^2+R^2(t,r)(d\theta^2
  +\sin^2\theta d\phi^2).
  \label{eq:lineelement}
\end{equation}
Using the comoving coordinates,
the stress-energy tensor $T^{\mu}_{~\nu}$ 
with vanishing radial pressure
is of the following form:
\begin{equation}
  T^{\mu}_{~\nu}= \pmatrix{
    -\epsilon & 0 & 0 & 0 \cr
    0 & 0 & 0 & 0 \cr
    0 & 0 & \Pi & 0 \cr
    0 & 0 & 0 & \Pi \cr
    },
\end{equation}
where $\epsilon(t,r)$, $\Pi(t,r)$ are
the energy density and
the tangential pressure, respectively.
From the Einstein equation and the equation of 
motion for the matter, we obtain
\begin{eqnarray}
  m&=&F,
  \label{eq:mconserve}\\
  \epsilon&=&\frac{F^{\prime}}{4\pi R^2 R^{\prime}}, 
  \label{eq:energydensitycomoving}\\
  e^{2\lambda}&=&R^{\prime 2}h^2, 
  \label{eq:grr} \\
  \nu^{\prime}&=&-\frac{1}{h}h_{,R}R^{\prime},
  \label{eq:lapse}\\
  \dot{R}^2 e^{-2\nu}&=&-1+\frac{2F}{R}+\frac{1}{h^2},
  \label{eq:energyofparticle}
\end{eqnarray}
where an arbitrary function $F=F(r)$ is the conserved
Misner-Sharp mass (Misner and Sharp 1973).
The dot and prime denote the partial derivatives
with respect to $t$ and $r$, respectively.
We have introduced a function $h=h(r,R)\ge 0$ as
\begin{equation}
  \Pi=-\frac{R}{2h}h_{,R}\epsilon,
  \label{eq:eos}
\end{equation}
where the comma denotes the partial derivative.
We should note that the definition of $h$ is slightly 
different from Magli (1997, 1998)'s notation. 
The dust limit is given by $h=h(r)$.
Eqs.~(\ref{eq:lapse}) and (\ref{eq:energyofparticle})
are coupled and cannot be integrated explicitly.

We can express the metric functions explicitly
by introducing the mass-area coordinate system
\begin{equation}
  ds^2 = -A(m,R) dm^2- 2B(m,R) dmdR -C(m,R) dR^2 
  +R^2 (d\theta^2 +\sin^2\theta d\phi^2),
\end{equation}
which was introduced by Ori (1990).
Since the derivation of the explicit solution 
was described in Magli (1998),
here we only present the results:
\begin{eqnarray}
  A&=&H\left(1-\frac{2m}{R}\right), \\
  B&=&-\frac{\sqrt{H}}{h|u|}, \\
  C&=&\frac{1}{u^2}, 
\end{eqnarray}
where
\begin{eqnarray}
  \sqrt{H}&=&\frac{R^{0}_{,m}h(m,R^{0}(m))}{|u^{0}|}
  +\int^{R}_{R^{0}}\frac{h}{x}\left(1+\frac{x}{2}\left(\frac{1}
      {h^2}\right)_{,m}\right)\left(-1+\frac{2m}{x}+\frac{1}{h^2}
  \right)^{-3/2}dx, 
  \label{eq:rootH}\\  
  u&\equiv&\frac{dR}{d\tau}=\pm\sqrt{-1+\frac{2m}{R}+\frac{1}{h^2}}, 
  \label{eq:velocityofshell}\\
  u^{0}(m)&\equiv&\pm\sqrt{-1+\frac{2m}{R^{0}(m)}
    +\frac{1}{h^2(m,R^{0}(m))}}, \\
  \label{eq:energydensityma}
  R^{0}(m)&\equiv& R(0,F^{-1}(m)),
\end{eqnarray}
the energy density is given as
\begin{equation}
  \epsilon=\frac{h}{4\pi R^2|u|\sqrt{H}},
\end{equation}
and we have assumed $\epsilon\ge 0$.

The shell-crossing singularity is the one 
characterized by $R^{\prime}=0$
and $R>0$, while the shell-focusing singularity 
is the one characterized by $R=0$.
Newman (1986) showed that the shell-crossing singularities
do not satisfy even LFC for a null geodesic.
Christodoulou (1984) showed that noncentral ($r>0$ or $m>0$)
shell-focusing singularities are not naked.
Therefore we concentrate on central ($r=0$ or $m=0$)
shell-focusing singularities.

If and only if the singularity is naked,
there exists an outgoing null
geodesic which emanates from the singularity.
In the mass-area coordinates, we can derive the root equation 
which probes the existence of such a geodesic as follows.
The radial null rays are determined by the equation
\begin{equation}
  \frac{dR}{dm}=\frac{-B\mp\sqrt{H}}{C}
  =\sqrt{H}|u|\left(\frac{1}{h}\mp|u|\right).
  \label{eq:drdm}
\end{equation}
We should note that the upper sign refers to an outgoing null ray
in a collapsing phase and an ingoing null ray in an expanding 
phase {\em at the same time}.
Similarly, the lower sign refers to an ingoing null ray
in a collapsing phase and an outgoing null ray
in an expanding phase {\em at the same time}.
Hereafter we mainly concentrate on 
a collapsing phase.

Here we define
\begin{equation}
  y\equiv\frac{R}{2m^{\beta}},
\end{equation}
where $\beta$ is determined by requiring that 
$y$ has a positive finite limit $y_0$ along the null geodesic.
Then, the regular center corresponds to $\beta\le 1/3$.
If we assume that the energy density at the center is positive, 
the regular center corresponds to $\beta= 1/3$.
Note that we will only consider naked singularities with 
such $\beta>1/3$.
It is noted that
we will assume the existence of every limit through this paper
in a sense including $\pm \infty$.
Then, from the l'Hospital's rule,
\begin{equation}
  y_{0}=\lim_{m\to0}\frac{R}{2m^{\beta}}
  =\lim_{m\to0}\frac{m^{1-\beta}}{2\beta}\frac{dR}{dm}
  =\left.\lim_{m\to0}\frac{m^{1-\beta}}{2\beta}\sqrt{H}|u|
  \left(\frac{1}{h}\mp |u|\right)\right|_{R=2y_0m^{\beta}}.
\label{eq:rooteqma}
\end{equation}
Therefore, we obtain the root equation
for the existence of the null geodesic from the central singularity 
\begin{equation}
  y_0 =\frac{1}{2\beta}\lim_{m\to0}
  \left[m^{3(1-\beta)/2}\sqrt{H}
  \sqrt{\left(-1+\frac{1}{h^2}\right)m^{-(1-\beta)}+\frac{1}{y_0}}
  \left(\frac{1}{h}\mp \sqrt{\frac{m^{1-\beta}}{y_0}-1+\frac{1}{h^2}}
    \right)\right],
  \label{eq:rooteq}
\end{equation}
where the limit is taken along $R=2y_{0}m^{\beta}$.
This equation was first derived by Magli (1998).
As seen in this equation, 
the existence of a future-directed 
outgoing null ray from the singularity
in a collapsing phase requires
\begin{equation}
  \frac{1}{3}< \beta\le 1.
\end{equation}
From $u^2\ge0 $ and $0<y_{0}<\infty$,
\begin{equation}
  \lim_{m\to 0}h\le 
  \left\{\begin{array}{ll}
        1&\mbox{for $\beta<1$} \\
        (1-y^{-1}_{0})^{-1/2}&\mbox{for $\beta=1$}
        \end{array}\right.,
\end{equation}
where the limit is taken
along the null ray which emanates from the singularity.
Here we should note that, if the singularity is 
{\em critically naked}, i.e., if 
\begin{equation}
  \lim_{m\to 0}\frac{2m}{R}=1
\end{equation}
holds
along the null ray,
higher order analysis is needed.
\section{curvature condition along null geodesic}
We consider a radial null geodesic
which emanates from or terminates at the naked singularity.
We prepare a parallely propagated 
tetrad $E_{i}:(i=1,2,3,4)$
with $E_{1}\cdot E_{1}=E_{2}\cdot E_{2}
=-E_{3}\cdot E_{4}=-E_{4}\cdot E_{3}=1$,
all other products vanish and $E_{4}$
is equal to the tangent vector $k^{\mu}$ of the null geodesic.
In a spherically symmetric space-time,
for the tetrad components of the Weyl tensor, 
\begin{equation}
  C^{m}_{~4n4}=0
\end{equation}
holds for $m,n=1,2$.
Define
\begin{equation}
  p\equiv\lim_{\lambda\to +0}\lambda^{\alpha} R_{44},
  \label{eq:defp}
\end{equation}
where $R_{44}$ is defined by
\begin{equation}
  R_{44}\equiv R_{\mu\nu}k^{\mu}k^{\nu}
  \label{eq:defR44}
\end{equation}
and $\lambda$ is the affine parameter such that
$\lambda\to +0$ corresponds to an approach to the singularity.
Then, from Clarke and Kr\'olak (1985) and Clarke (1993),
\begin{lm}
For the radial null geodesic which emanates from 
or terminates at the singularity 
in the spherically symmetric space-time:
SCC is satisfied if $p$ is positive for $\alpha=2$,
and not satisfied if $p$ is equal to $0$ for $\alpha<2$;
LFC is satisfied if $p$ is positive for $\alpha=1$,
and not satisfied if $p$ is equal to $0$ for $\alpha<1$.
\end{lm}

Since the null geodesic is given as
\begin{equation}
  k^{R}=\frac{-B \mp \sqrt{H}}{C}k^{m},
  \label{eq:nullcondition}
\end{equation}
we obtain, 
from the form of the stress-energy tensor,
\begin{equation}
  R_{44}=8\pi \epsilon u^2 H (k^{m})^{2}.
\end{equation}
Then, 
\begin{equation}
  \lambda^2 R_{44}
  = \frac{1}{2} 
  |u|h\sqrt{H}
  \left(\frac{2m}{R}\right)^2
  \left(\frac{\lambda}{m}\frac{d m}{d\lambda}
  \right)^2 
  \approx\frac{\beta q^2}{y_{0}}
  m^{1-\beta}\frac{h^2}{1\mp h|u|}
\end{equation}
holds, where $q$ is defined as 
\begin{equation}
  q\equiv \lim_{m\to 0}
  \frac{d\ln m}{d\ln\lambda}
\end{equation}
and we have used Eq.~(\ref{eq:rooteqma}).

We should note that, for $(\beta,y_{0})\ne (1,1)$,
\begin{equation}
  0<\lim_{m\to 0}\frac{h^2}{1-h|u|}<\infty
\end{equation}
and 
\begin{equation}
  0\le\lim_{m\to 0}\frac{h^2}{1+h|u|}<\infty
\end{equation}
hold,
where the equality holds only when
\begin{equation}
  \lim_{m\to 0}h=0
\end{equation}
is satisfied.
Therefore, 
\begin{equation}
  R_{44}\propto \lambda^{q(1-\beta)-2}
  \label{eq:R44}
\end{equation}
holds
for the outgoing null geodesic with $0<q<\infty$ and $ (\beta,y_{0})\ne
(1,1)$, and for the ingoing null geodesic with
$0<q<\infty$, $ (\beta,y_{0})\ne(1,1)$ and $\lim_{m\to 0}h\ne 0$.
In summary, we present the following theorems:
\begin{th}
For the outgoing radial null geodesic which emanates from the 
noncritically naked singularity with $0<q<\infty$:
if and only if $1/3<\beta<1$ and 
$(1-\beta)^{-1}<q<\infty$ are satisfied, neither SCC nor LFC holds, 
if and only if $1/3<\beta<1$ and $0<q\le (1-\beta)^{-1}$ are satisfied, 
not SCC but only LFC
holds,
and if and only if $\beta=1$ is satisfied, both SCC and LFC hold.  
\end{th}
\begin{th}
For the ingoing radial null geodesic which terminates at the 
noncritically naked singularity
with $0<q<\infty$ and $lim_{m\to 0}h\ne 0$:
if and only if $1/3<\beta<1$ and 
$(1-\beta)^{-1}<q<\infty$ are satisfied, neither SCC nor LFC holds, 
if $1/3<\beta<1$ and $0<q\le (1-\beta)^{-1}$ are satisfied, 
not SCC but only LFC holds,
and if and only if $\beta=1$ is satisfied, both SCC and LFC hold.  
\end{th}

In order to estimate $q$,
we must solve the null geodesic equation
\begin{equation}
  \frac{d}{d\lambda}\left(\sqrt{H}k^{m}\right)
  \pm \frac{1}{2}\left[A_{,R}+2 B_{,R} |u|
    \sqrt{H}\left(\frac{1}{h}\mp |u|\right)
    +C_{,R} u^2 H \left(\frac{1}{h}\mp |u|\right)^2
  \right](k^{m})^{2}=0,
  \label{eq:nullgeodesiceqma}
\end{equation}
where we have used the 
null condition (\ref{eq:nullcondition}).
From Eq.~(\ref{eq:rootH}), we obtain
\begin{equation}
  \sqrt{H}_{,R}=\frac{h}{2|u|^3}\left(\frac{2}{R}
    +\left(\frac{1}{h^2}\right)_{,m}\right).
\end{equation}
Using this, the following expressions are derived.
\begin{eqnarray}
  A_{,R}&=&\frac{2m}{R^2}H+\frac{\sqrt{H}}{|u|^3}h
  \left(1-\frac{2m}{R}\right)
  \left(\frac{2}{R}+\left(\frac{1}{h^2}
    \right)_{,m}\right), \\
  B_{,R}&=&-\frac{1}{2|u|^4}\left(\frac{2}{R}+\left(\frac{1}{h^2}
      \right)_{,m}\right)
    -\frac{\sqrt{H}}{|u|^3}\frac{1}{h}
    \left[\left(1-\frac{2m}{R}\right)\frac{h_{,R}}{h}+
      \frac{m}{R^2}\right],\\
  C_{,R}&=&\frac{2}{u^4}\left(\frac{m}{R^2}+\frac{1}{h^2}
    \frac{h_{,R}}{h}\right).
\end{eqnarray}  
Using these expressions, we finally obtain the radial 
null geodesic equation for $m\to 0$
in the explicit form
\begin{equation}
  \frac{d^2m}{d\lambda^2}=\left[1-\beta
    -\frac{1}{2}\frac{h^2}{1\mp h|u|}
    \left(\frac{2m}{R}+\frac{d}{d\ln m}\frac{1}{h^2}\right)\right]
      \frac{1}{m}\left(\frac{dm}{d\lambda}\right)^2,
      \label{eq:nullgeo}
\end{equation}
where the ordinary derivative is taken along $R=2y_{0}m^{\beta}$.
In evaluating the right hand side 
of Eq.~(\ref{eq:nullgeo}), we have used
\begin{eqnarray}
  R&\approx&2y_{0}m^{\beta}, 
  \label{eq:R}\\
  |u|&\approx&m^{(1-\beta)/2}\sqrt{\frac{1}{y_{0}}
    +m^{-(1-\beta)}\left(\frac{1}{h^2}-1\right)},
  \label{eq:u}\\
  \sqrt{H}&\approx&\beta\frac{R}{m}\frac{h}{|u|(1\mp h|u|)}.
  \label{eq:rtH}
\end{eqnarray}
On the other hand, if $m$ is proportional to $\lambda^{q}$
along the null geodesic, the following equation holds:
\begin{equation}
  \frac{d^2m}{d\lambda^2}=\left(1-\frac{1}{q}\right)\frac{1}{m}
  \left(\frac{dm}{d\lambda}\right)^2.
  \label{eq:mlambdaq}
\end{equation}
Comparing Eqs.~(\ref{eq:nullgeo}) and (\ref{eq:mlambdaq}), 
we can determine $q$. 
Therefore the curvature divergence 
along the null geodesic 
is determined 
only from $\beta$ and $h$ along the geodesic.

\section{gravity dominance condition}
\label{sec:gravitydom}
Here we define the gravity dominance condition (GDC)
and the gravity-dominated singularity as
follows:
\newtheorem{df}{Definition}
\begin{df}[Gravity Dominance Condition]
For the geodesic which emanates from or terminates at the singularity,
we have
\begin{equation}
  \lim_{m\to 0}\frac{R}{2m}\left(\frac{1}{h^2}-1\right)= 0.
\label{eq:condition1}
\end{equation}
\end{df}
\begin{df}[Gravity-Dominated Singularity]
A singularity
is said to be gravity-dominated if and only if 
GDC is satisfied
for every causal geodesic which emanates from or terminates at 
the singularity.
\end{df}
GDC is satisfied for the geodesic which 
emanates from or terminates at a 
very wide class of naked singularities.
Furthermore, if the gravitational collapse of 
physical matter from regular initial data
results in the central naked singularity formation,
GDC is satisfied for the null geodesic, 
at least within our knowledge.
The important example is the central singularity 
in the collapse of the spherical cluster of
counterrotating particles which will be discussed 
in Sec. VI.

If GDC is satisfied for the null geodesic,
the collapse is induced dominantly by the gravitational 
potential
(see Eq.~(\ref{eq:energyofparticle}) 
or (\ref{eq:velocityofshell}))
and the null geodesic equation
is controlled only by the gravitational potential
(see Eq.~(\ref{eq:nullgeo}).
The latter can be shown by the following proposition.
\begin{pr}
If GDC is satisfied, then 
\begin{equation}
  \lim_{m\to 0}\frac{R}{2m}\frac{d}{d\ln m}\frac{1}{h^2}=0
\label{eq:condition2}
\end{equation}
holds.
\end{pr}
{\em Proof.} We use $R\approx 2y_{0}m^{\beta}$ along the null geodesic.
Then, for $\beta<1$, the l'Hospital's rule applies because we have
assumed the exsitence of the limit.
From condition (\ref{eq:condition1}), the proposition holds.
For $\beta=1$, we set $f \equiv h^{-2}-1$.
Then condition (\ref{eq:condition1}) implies $\lim_{x\to 0}f(x)=0$.
From the mean value theorem, 
there exists $c\in (0,x)$ for any $x>0$ such that
\[ \left|c f^{\prime}(c)\right|
=\left|c\frac{f(x)}{x}\right|\le|f(x)|. \]
Because we have assumed the existence of the limit,
it must be zero.\hfill $\Box$

If GDC is satisfied for the null geodesic, 
Eqs.~(\ref{eq:u})
and (\ref{eq:rtH}) become
\begin{equation}
  |u|\approx y_{0}^{-1/2}m^{(1-\beta)/2},
\end{equation}
and
\begin{equation}
  \sqrt{H}\approx\left\{
  \begin{array}{ll}
    2\beta y_{0}^{3/2}m^{-3(1-\beta)/2},& \mbox{for $\beta<1$}\\
    \displaystyle\frac{2 y_{0}^{2}}{y_{0}^{1/2}\mp 1}, 
    & \mbox{for $\beta=1$ and
    $y_{0}\ne1$}\\
  \end{array}\right.. 
\end{equation}

For $\beta<1$, since Eq.~(\ref{eq:nullgeo}) becomes
\begin{equation}
  \frac{d^2m}{d\lambda^2}
    =(1-\beta)\frac{1}{m}\left(\frac{dm}{d\lambda}\right)^2,
\end{equation}
we obtain
\begin{equation}
  q=\frac{1}{\beta}.
\end{equation}
Then, $1<q<3$ and $R\propto \lambda$ hold.
Eq.~(\ref{eq:R44}) becomes
\begin{equation}
  R_{44}\propto \lambda^{-3+\beta^{-1}}.
\end{equation}
Therefore SCC is not satisfied.
LFC is satisfied for $1/2\le \beta <1$, while
LFC is not satisfied for $1/3<\beta<1/2$.

For $\beta=1$ and $y_{0}\ne 1$, Eq.~(\ref{eq:nullgeo})
becomes
\begin{equation}
  \frac{d^2m}{d\lambda^2}
  =-\frac{1}{2(y_{0}^{1/2}\mp 1)y_{0}^{1/2}}
  \frac{1}{m}\left(\frac{dm}{d\lambda}\right)^2.
\end{equation}
Therefore we obtain
\begin{equation}
  q=\frac{2(y_{0}^{1/2}\mp 1)y_{0}^{1/2}}
  {2(y_{0}^{1/2}\mp 1)y_{0}^{1/2}+1},
\end{equation}
Then, $0<q<1$ and $R\propto \lambda^{q}$ hold.
Eq.~(\ref{eq:R44}) becomes
\begin{equation}
  R_{44}\propto \lambda^{-2}.
\end{equation}
Therefore both SCC and LFC are satisfied.

In summary, we present the following theorems:
\begin{th}
Suppose that GDC is satisfied for a radil null geodesic
which emanates from or terminates at 
the noncritically naked singularity.
If and only if $1/3<\beta<1/2$ is satisfied, neither SCC nor LFC holds,
if and only if $1/2\le \beta<1$ is satisfied, not SCC but only LFC holds,
and if and only if $\beta=1$ is satisfied, then both SCC and LFC hold,
for the radial null geodesic which emanates from or terminates 
at the singularity.
\end{th}
\begin{th}
Suppose that GDC is satisfied for a radil null geodesic
which emanates from or terminates at 
the noncritically naked singularity.
Along the radial null geodesic,
\[ \lim_{m\to 0}\frac{R}{\lambda} \]
is nonzero finite value or positive infinity. 
If and only if the limit converges, SCC does not hold,
and if and only if the limit diverges, both SCC and LFC 
hold, for the null geodesic.
\end{th}

\section{curvature condition along timelike geodesic}
Here we consider a timelike geodesic which
terminates at the singularity.
We prepare a parallely propagated tetrad $E_{i}$ with
$E_{1}\cdot E_{1}=E_{2}\cdot E_{2}=E_{3}\cdot E_{3}=-E_{4}\cdot E_{4}=1$,
all other products vanish and $E_{4}$ is equal to the tangent vector $k^{\mu}$
of the timelike geodesic.
We can define $p$ and $R_{44}$ by Eqs.~(\ref{eq:defp})
and (\ref{eq:defR44}), respectively, 
also for the timelike geodesic.
From Clarke and Kr\'olak (1985) and Clarke (1993),
the following lemma holds.

\begin{lm}
For the timelike and null geodesic 
which emanate from
or terminate at the singularity:
SCC
is satisfied if 
$p$ is positive for $\alpha=2$;
LFC
is satisfied 
if $p$ is positive for $\alpha=1$.
\end{lm}

It seems to be cumbersome to 
examine the curvauture diveregnce
along
all possible timelike geodesics.
Then, we consider the simplest timelike geodesic, i.e.,
$r=0$.
It is easy to find that $r=0$ is a timelike geodesic
when the center is regular.
As a matter of convenience, 
we adopt the coordinate system (\ref{eq:lineelement}).
Along $r=0$, the $R_{44}$ is calculated as
\begin{equation}
  R_{44}=4\pi \epsilon = \frac{F^{\prime}}{R^2R^{\prime}},
  \label{eq:R44timelike}
\end{equation}
where we have used $\Pi=0$ at the regular center 
which will be seen in Sec. VII.

We will consider the situation in which the central
singularity occurs at $t=0$ 
from the regular initial data at $t=t_{0}<0$.
We choose the radial coordinate $r$ as $r=R(t_{0},r)$.
From regularity of the center, we obtain, 
for $t_{0}\le t< 0$, 
\begin{eqnarray}
  F(r)&=&F_{3}r^3+\cdots,\\
  R(t,r)&=&R_{1}(t)r+\cdots, \\
  \nu(t,r)&=&\nu_{0}(t)+\cdots, 
\end{eqnarray}
where ``$\cdots$'' means the higher order terms 
with respect to $r$.
As we assume the positivity of the energy density at the center
at $t=t_{0}$, $F_{3}>0$ must be satisfied.
We set $\nu_{0}(t)=0$ by using the scaling freedom of time
coordinate. From this choice, the time coordinate $t$
can coincide with the proper time $\tau$ at the center.
Substituting into Eq.~(\ref{eq:R44timelike}), 
the value of $R_{44}$ at the center is 
written as
\begin{equation}
  R_{44}=\frac{3 F_{3}}{R_{1}^3}.
  \label{eq:R44center}
\end{equation}
Note that $R_{1}=0$ corresponds to the occurrence of 
the central singularity.

From the equation of motion of each mass shell 
(\ref{eq:energyofparticle}),
it is required that
\begin{equation}
  \frac{1}{h^2}-1=h_{1}(t)r^2+\cdots.
\end{equation}
The lowest order of Eq.~(\ref{eq:energyofparticle})
becomes
\begin{equation}
  \dot{R_{1}}^2=\frac{2F_{3}}{R_{1}}+h_{1}.
  \label{eq:R1}
\end{equation}
Here we assume that GDC
is satisfied for the 
timelike geodesic $r=0$, 
where it should be noted that the value of 
\begin{equation}
  \frac{R}{2m}\left(\frac{1}{h^2}-1\right)
\end{equation}
at $r=0$ is understood as the limit of $r\to 0$.
Then it is found that
\begin{equation}
  \lim_{t \to 0}\frac{R_{1}h_{1}}{F_{3}} =0.
\end{equation}
Hence, Eq.~(\ref{eq:R1}) becomes
\begin{equation}
  \dot{R_{1}}^2\approx \frac{2F_{3}}{R_1}
\end{equation}
in the limit of $t\to 0$.
This is integrable as
\begin{equation}
  R_{1}\approx \left(\frac{9F_{3}}{2}\right)^{1/3}(-t)^{2/3}
  =\left(\frac{9F_{3}}{2}\right)^{1/3}{(-\tau)^{2/3}}.
\end{equation}
Eq.~(\ref{eq:R44center}) becomes
\begin{equation}
  R_{44}\approx \frac{2}{3}\frac{1}{(-\tau)^2}.
\end{equation}
Therefore, for the timelike geodesic $r=0$, 
both SCC and LFC are satisfied.
\begin{th}
  If GDC is satisfied for the timelike geodesic $r=0$
which terminates at the singularity, then
both SCC and LFC are satisfied for the timelike geodesic.
\end{th}

\section{examples}
\subsection{dust collapse}
The spherically symmetric dust collapse has been
analyzed in the context of naked singularities
by Eardley and Smarr (1979), 
Christodoulou (1984), Newman (1986), Joshi and Dwivedi (1993),
Singh and Joshi (1996), and Jhingan, Joshi and Singh (1996).
The stability of the Cauchy horizon 
against nonspherical perturbation
was recently discussed by Iguchi, Nakao and Harada (1998)
and Iguchi, Harada and Nakao (1998).
 
The dust fluid is given by $h(r,R)=h(r)$.
For simplicity we restrict our attention to the marginally
bound collapse which is given by $h=1$.
It is trivial that the singularity is gravity-dominated.
The space-time is given by the LTB solution.
The solution in the mass-area coordinates is given 
by Ori (1990) and Magli (1998).
The solution contains an arbitrary function $F(r)$.
Here we choose the comoving radial coordinate $r$ as
$ r=R(t=t_{0},r)$, i.e., $R^{0}(m)=F^{-1}(m)$.

First we give the function $F(r)$ as
\begin{equation}
  F(r)=F_{3}r^3+F_{5}r^5+\cdots,
\label{eq:smooth}
\end{equation}
which corresponds to generic smooth initial data.
For $F_3>0$ and $F_5<0$, 
Eq.~(\ref{eq:rooteq}) has a finite positive root 
\begin{equation}
y_{0}=\left(\frac{-F_5}{4\sqrt{2}F_3^{13/6}}\right)^{2/3}
\end{equation}  
with $\beta=7/9$.
From the results in Sec.~\ref{sec:gravitydom},
not SCC but only LFC is satisfied for the radial null geodesic
which emanates from the singularity.

Next, if we give $F(r)$ as
\begin{equation}
  F(r)=F_{3}r^3+F_{6}r^6+\cdots,
\end{equation}
which corresponds to nongeneric regular initial data.
For $F_{3}>0$ and $F_{6}<-(26\sqrt{2}+15\sqrt{6})F_{3}^{5/2}$,
Eq.~(\ref{eq:rooteq}) has a finite positive root $y_{0}$
with $\beta=1$, where $y_{0}> 1$ is expressed using the root
of some quartic equation.
Then, from the results in Sec.~\ref{sec:gravitydom},
both SCC and LFC are satisfied for the outgoing radial null
geodesic which emanates from the singularity.

For the above two cases,
the curvature strength is exactly the
same for the ingoing radial null geodesic which terminates at the  
singularity. 

On the other hand, both SCC and LFC are satisfied 
for the timelike geodesic 
$r=0$.
This fact was already shown by Deshingkar, Joshi and Dwivedi (1999).
This can be confirmed by the result of Sec. V since GDC
is also satisfied for the timelike geodesic.
This is the case not only for marginally bound collapse but also
for nonmarginally bound collapse because GDC
is satisfied for the timelike geodesic.
\subsection{cluster of counterrotating particles}
The dynamical spherical cluster of counterrotating particles
was introduced and analyzed by Datta (1970), Bondi (1971)
and Evans (1976).
The explicit solution for the metric functions
was derived by Harada, Iguchi and Nakao (1998).
They also examined the occurrence of naked singularity.

We again restrict our attention to the marginally bound collapse.
Then, the model is given by
\begin{equation}
  h^2=1+\frac{L^2}{R^2},
\end{equation}
where $L=L(m)$ is the specific angular momentum.
We give $F(r)$ as in Eq.~(\ref{eq:smooth}).
If $L(m)$ is given by $L=4m$, 
the metric functions are expressed by elementary functions.
For this case, Harada, Iguchi and Nakao (1998) showed that
Eq.~(\ref{eq:rooteq}) has a finite positive root 
\begin{equation}
y_{0}=\left(\frac{24F_3^2-F_5}{4\sqrt{2} F_3^{13/6}}\right)^{2/3},
\end{equation}
for $F_5 < 24 F_3^2$ with $\beta=7/9$.
Note that $F_5 < 24 F_3^2$ is the same as the requirement of 
no shell-crossing singularity.
GDC is satisfied for the null geodesic.
From the results in Sec.~\ref{sec:gravitydom},
not SCC but only LFC is satisfied for the radial null geodesic
which emanates from or terminates at the singularity.

On the other hand, it is found that GDC is also satisfied
for the timelike geodesic $r=0$.
From the result of Sec. V, 
both SCC and LFC are satisfied for this timelike geodesic.
This is the case not only for marginally bound 
collapse but also for nonmarginally bound collapse because GDC
is satisfied for the timelike geodesic.
\subsection{\protect$\Pi=k\epsilon$}
We consider the equation of state
\begin{equation}
  \Pi=k\epsilon,
\end{equation}
where $k$ is a constant.
This will be the simplest nontrivial equation of state
for tangential pressure.
Singh and Witten (1997) examined the motion of a fluid
with this equation of state.
From Eqs.~(\ref{eq:lapse}) and (\ref{eq:eos}),
\begin{equation}
  \nu^{\prime}=2k\frac{R^{\prime}}{R}
\end{equation}
holds.
Since regularity requires
\begin{eqnarray}
  \nu&=&\nu_{0}(t)+O(r^2), \\
  R&=&R_{1}(t)r+O(r^3),
\end{eqnarray}
it is impossible to set regular initial data
for $k\ne 0$.
Therefore this model is not appropriate 
for a probe of CCH.

\section{concluding remarks}
The nakedness and curvature strength 
of shell-focusing singularity
in the spherically symmetric collapse of a fluid
with vanishing radial pressure has been investigated.
Along the first radial null ray from the naked singularity,
$R\approx 2y_{0} m^{\beta}$ $(1/3<\beta\le 1)$
is satisfied.
The $y_{0}$ and $\beta$ are determined by some root equation.
The $\beta$ is closely related to the curvature strength 
of the singularity for the null geodesic which
emanates from or terminates at the singularity.
Roughly speaking, 
$\beta=1$ means SCC and vice versa.

Then, we have defined GDC 
for the geodesic which emanates from or terminates at
the singularity.
Suppose that GDC is satisfied for the null geodesic. 
For this class of noncritically naked singularities, 
if and only if 
$1/3<\beta<1/2$ is satisfied, neither SCC nor LFC holds,
if and only if 
$1/2\le \beta<1$ is satisfied, not SCC but only LFC holds,
and if and only if $\beta=1$ is satisfied, 
both SCC and LFC hold.
Furthermore, for this class of noncritically naked singularities,
if and only if $\lim_{m\to 0}\lambda^{-1}R$ diverges, 
SCC is satisfied.

We also have examined whether or not the curvature 
divergence condition 
is satisfied for a timelike geodesic.
Suppose that GDC is satisfied for the timelike geodesic $r=0$
which terminates at the singularity.
Then, we have found that both SCC and LFC are satisfied 
for the timelike geodesic.

We have applied this formalism to the dust collapse and 
the collapse of counterrotating particles.
It is noted that, with vanishing radial pressure,
only if the ratio of the tangential pressure
to the energy density vanishes at the center,
it is possible to set regular initial data
which is important ingredient when we consider physical situations.

Even if we include the tangential pressure,
nakedness and curvature strength of the singularity
are very similar to those of the dust model if the singularity
is gravity-dominated.
On the other hand, if the singularity is not gravity-dominated,
then we may expect that the tangential pressure plays 
a crucial role
in the nakedness of the singularity and the extendibility
of the space-time beyond the singularity.

\acknowledgments
We are grateful to T. Nakamura and M. Siino
for helpful discussions.
We are also grateful to H. Sato
for his continuous encouragement.
This work was partly supported by the 
Grant-in-Aid for Scientific Research (No. 9204)
and for Creative Basic Research (No. 09NP0801)
from the Japanese Ministry of
Education, Science, Sports and Culture.
\appendix

\end{document}